%% file: ANA-FTAG-2023-05-PAPER.tex
\documentclass[PAPER, atlasdraft=false, coverpage=false, cernpreprint=true, UKenglish, texmf, orcidlogo]{atlasdoc}
\usepackage[subcaption]{atlaspackage}
\usepackage{multirow}
\usepackage{comment}
\usepackage{atlasbiblatex}

\usepackage[misc]{atlasphysics}

\usepackage{siunitx}

\graphicspath{{logos/}{figures/}}

\usepackage{ANA-FTAG-2023-05-PAPER-defs}

\AtlasTitle{Transforming jet flavour tagging at ATLAS}

\AtlasAbstract{%

Jet flavour tagging enables the identification of jets originating from heavy-flavour quarks in proton--proton collisions at the Large Hadron Collider, playing a critical role in its physics programmes. This paper presents GN2, a transformer-based flavour tagging algorithm deployed by the ATLAS Collaboration that represents a different methodology compared to previous approaches. Designed to classify jets based on the flavour of their constituent particles, GN2 processes low-level tracking information in an end-to-end architecture and incorporates physics-informed auxiliary training objectives to enhance both interpretability and performance. Its performance is validated in both simulation and collision data. The measured \cjet (\ljet) rejection in data is improved by a factor of 3.5 (1.8) for a 70\% \bjet tagging efficiency, compared to the previous algorithm. GN2 provides substantial benefits for physics analyses involving heavy-flavour jets, such as measurements of Higgs boson pair production and the couplings of bottom and charm quarks to the Higgs boson, and demonstrates the impact of advanced machine learning methods in experimental particle physics.
}

\author{The ATLAS Collaboration}

\AtlasRefCode{FTAG-2023-05}

\PreprintIdNumber{CERN-EP-2025-103}

\AtlasJournal{Nature Communications}
\AtlasJournalRef{Nature Commun. 17 (2026) 541}
\AtlasDOI{10.1038/s41467-025-65059-6}

\hypersetup{pdftitle={ATLAS document},pdfauthor={The ATLAS Collaboration}}

\begin{document}

\maketitle

%

%
%

%
%
%
%

% The next lines are included from the .//sections/introduction.tex input file
%
%
%
%
%
%
%
%
%

The Large Hadron Collider (LHC)~\cite{Evans:2008zzb} is the world's most
powerful particle collider. It is used to extend the boundaries of our
understanding of fundamental particles and their interactions. It offers a
unique opportunity to test the Standard Model (SM) of particle physics, as well
as search for new phenomena beyond the Standard Model (BSM). The demanding
experimental conditions at the LHC necessitate continuous innovation by the
main experiments, pushing them to apply cutting-edge technologies to
efficiently identify physics processes of interest within the largest
proton--proton ($pp$) collision dataset ever recorded. Hadronic jets,
collimated streams of particles initialised by quarks or gluons, are the most
abundant physics objects in $pp$ collision events, and their characteristics
are widely utilised in data analyses.

The flavour of a hadronic jet is determined by the types of hadrons or leptons
it contains. Flavour tagging concerns the classification of hadronic jets into
those containing \bhadrons (\bjets), \chadrons (\cjets), hadronic
\(\taulep\) decays (\taujets), and none of the above (\ujets), using algorithms sensitive to the
distinctive properties of the respective classes. Since the beginning of
\RunOne of the LHC (2009--2013), the ATLAS experiment~\cite{PERF-2007-01,GENR-2019-02} has achieved continuous improvement in the performance of
these algorithms. The progress has mostly been
driven by the integration of machine-learning techniques, including boosted
decision trees and neural networks. The state-of-the-art algorithms used thus far to
analyse the data at \(\sqs = \SI{13}{\tera\electronvolt}\) from \RunTwo of the
LHC (2015--2018)~\cite{FTAG-2018-01,FTAG-2019-07} led to very impactful physics
results such as the observations of the Higgs boson decaying to bottom
quarks~\cite{ATLAS:2018kot} and its production in association with a pair of
top quarks~\cite{ATLAS:2018mme}. Flavour tagging plays an essential role in the
comprehensive research programme of ATLAS, which includes precision measurements
of the Higgs boson~\cite{ATLASHIGGS}, top quark~\cite{ATLASTOP} and other SM
processes~\cite{ATLASSM}, as well as the searches for
supersymmetry~\cite{ATLASSUSY} and other BSM phenomena~\cite{ATLASEXOTIC}. This
work describes a flavour tagging algorithm developed by the ATLAS
Collaboration for the analysis of data from \pp collisions recorded during \RunTwo (2015--2018) and
\RunThr (2022--2026) of the LHC at centre-of-mass energies of \(\sqs = \SI{13}{\tera\electronvolt}\) and \(\sqs = \SI{13.6}{\tera\electronvolt}\), respectively.

Flavour-tagging techniques rely on the long lifetime, high mass, high decay
multiplicity and characteristic decay modes of \(b\)- and \chadrons, and  the
properties of heavy-quark fragmentation~\cite{HFLAV:2022esi}. The typical
lifetime of the order of \(\tau \approx
\SI{1.5}{\pico\second}\)~\cite{HFLAV:2022esi,BaBar:2001mmd,ATLAS:2024B0Lifetime}
for \bhadrons in jets with transverse momenta in the range from tens to
hundreds of \si{\giga\electronvolt} results in them travelling a mean flight
length \( \langle l \rangle = \beta \gamma c \tau \) in the range from few
millimetres to centimetres before decaying, which often leads to a secondary
vertex significantly displaced from the collision point. Displaced vertices can
also be produced by \chadrons, which have lifetimes of \(\tau \approx
\SI{0.2}-\SI{1.0}{\pico\second}\), depending on the
species~\cite{,Belle-II:2021cxx,Belle-II:2022ggx,10.1093/ptep/ptaa104}, and
\(\tauleps\), which have a lifetime of \(\tau \approx \SI{0.29}{\pico\second}\)
but a much lower decay multiplicity~\cite{Belle:2013teo,10.1093/ptep/ptaa104}.
The majority of \bjets also contain a tertiary vertex from the decay of the
\chadron produced in the \bhadron decay.

The traditional flavour-tagging algorithms developed by the ATLAS Collaboration
are based on a two-stage approach~\cite{FTAGRun1,FTAG-2018-01,FTAG-2019-07}. In
the first step, specialised low-level algorithms employ complementary
approaches to extract information from the trajectories of the charged-particle
constituents (\enquote{tracks}) associated with the jet. These specialised
algorithms either rely on the properties of individual tracks or leverage their
correlations with properties of other tracks to explicitly reconstruct
displaced vertices. In the second step, the outputs of low-level algorithms are
subsequently combined in a high-level multivariate classifier to maximise
performance. The most recent algorithm employed by the ATLAS Collaboration,
following this paradigm, is a deep neural network (DL1d) that
leverages a low-level track-based algorithm
(DIPS)~\cite{ATL-PHYS-PUB-2020-014} based on Deep
Sets~\cite{Komiske_2019}. DL1d has already improved the performance by a factor
of 1.3 relative to the most advanced algorithm used in published Run-2 physics
analyses~\cite{FTAG-2019-07}.

The introduction of graph neural networks for object reconstruction in particle physics experiments~\cite{jonathan} prompted a shift in the design strategy of the ATLAS Collaboration. This led to the development of the General Network (GN) series of flavour-tagging algorithms, which directly process track and jet information and are trained using target labels extracted from Monte Carlo (MC) simulation. In parallel, the CMS Collaboration followed a similar trajectory, evolving from two-stage approaches~\cite{CMSBTagging,deepjet} to unified, end-to-end network architectures~\cite{ParticleNet,UParT,CMS-PAS-BTV-22-001}.

\begin{figure}[!t]
\centering
\includegraphics[width=\textwidth]{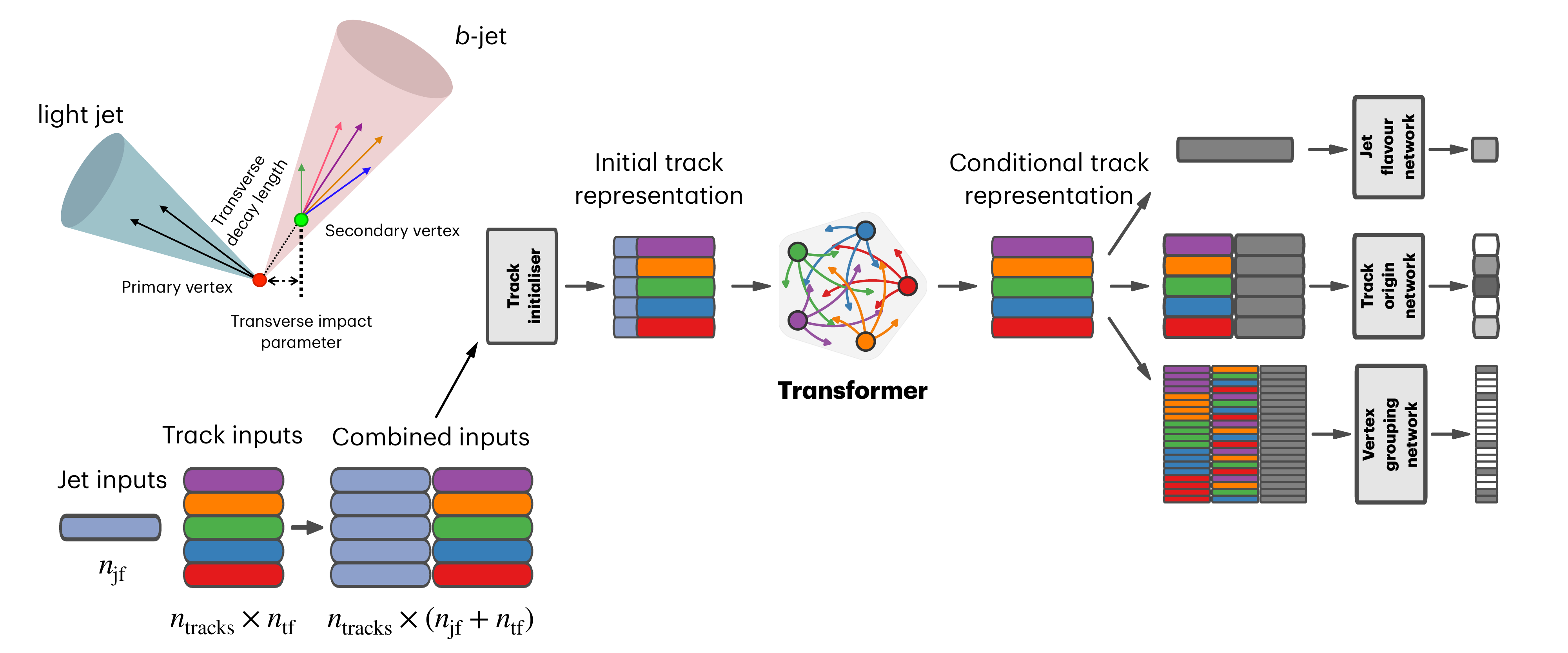}
\caption{
Illustration of the GN2 algorithm with jet and track input variables, discriminating between jet flavours by exploiting secondary vertices and other properties stemming from the displaced decays of $b$-hadrons, in the transverse plane. The jet features are copied for each track associated with the jet. The combined vectors are then fed into a per-track initialisation network, followed by a transformer encoder and a global representation of the jet. $n_{\mathrm{jf}}$ ($n_{\mathrm{tf}}$) corresponds to the number of jet (track) features. The pooled jet representation and output track embeddings are provided as inputs to the three task-specific networks. Details of the GN2 architecture are summarised in the Methods section.  }
\label{fig:gn2_sketch}
\end{figure}

The ATLAS GN tagger uses jet flavour prediction as its primary training target and introduces auxiliary training objectives to reconstruct the internal structure of a jet by grouping tracks originating from a common vertex and by predicting the underlying physics process from which each track originated. Such physics domain knowledge is embedded in a combined loss
function that enables a simultaneous optimisation, instead of relying on
manually optimised low-level algorithms.  This flexible structure allows the
swift re-tuning of the algorithms to suit alternative experimental conditions
or physics goals. A demonstrator version, GN1, achieves the above design goals
using a graph-neural-network~\cite{ATL-PHYS-PUB-2022-027}, while the deployment
version, GN2, applies a single transformer model~\cite{transformers},
illustrated in Figure~\ref{fig:gn2_sketch}.

Details of the algorithm
architectures are summarised in the Methods section, together with descriptions
of the ATLAS detector, simulation samples, physics objects, and analysis
strategies.

GN2 achieves a remarkable performance boost compared with the DL1d algorithm, with
improvements by a factor of 1.5–4 observed in its major experimental
applications. The deployment of GN2 should greatly enhance the physics reach of
ATLAS in flagship analyses, such as the search for Higgs pair production and the
$c$-quark Yukawa coupling measurement, for which the projected sensitivity at the High
Luminosity LHC is improved by up to 30\%~\cite{CMS:2025hfp}. These improvements
do not come with a strong dependence on the choice and configuration of the MC
event generator, and are confirmed by measured performance in recorded collisions.
The innovative auxiliary training objectives bring excellent interpretability
and opens up new avenues for future applications.

To facilitate future developments and strengthen the connections between
collider experiments and the broader scientific research community, a subset of
the training sample with all the required information to train GN2 can be
acquired via the CERN Open Data Portal~\cite{OpenData,jetset-examples}.

%
%
%

% End of text imported from the .//sections/introduction.tex input file

%
\section*{Results}
\label{sec:results}
%

% The next lines are included from the .//sections/results.tex input file
\textbf{Algorithm performance in simulation}

\noindent The performance of a \(b\)-tagging algorithm is evaluated based on
its ability to reject \(c\)-, \(\tau\)- and \ljets while maintaining a desired
\bjet tagging efficiency. Similarly, the \(c\)-tagging performance is assessed
by its capability to distinguish \(c\)-jets from the other jet flavours. The
data samples used for training and evaluation of the model must contain jets
from all flavour classes. This is achieved using jets sampled from a mixture
of simulated top quark pair (\ttbar) and \Zprime events, where the latter
sample considers a hypothetical heavy BSM particle,
\Zprime~\cite{Langacker:2008yv}, which can decay into pairs of $b$-quarks,
$c$-quarks, \(\tauleps\) or light quarks, to populate jets in the \TeV\ regime.
The samples are simulated with MC event generators at centre-of-mass energies
of both \(\sqs = \SI{13}{\tera\electronvolt}\) and \(\sqs
=\SI{13.6}{\tera\electronvolt}\). All simulated events are processed through the ATLAS detector simulation~\cite{SOFT-2010-01} based on
\GEANT~\cite{Agostinelli:2002hh,geant4_2006,geant4_2016}. Further details on
the simulation samples and the jet flavour labelling are discussed in the
Methods section. A mixture of samples generated at \(\sqs
=\SI{13.6}{\tera\electronvolt}\) and \(\sqs =\SI{13}{\tera\electronvolt}\) is used in the training, to achieve similar performance in both conditions. In this section, the performance evaluated with Run-3 samples at \(\sqs =\SI{13.6}{\tera\electronvolt}\) is presented.
\noindent Jets are classified for \(b\)-tagging using a single discriminant
\(D_b\), which combines the algorithm's jet flavour prediction output
probabilities of a jet being a \bjet (\(p_b\)), a \cjet (\(p_c\)), a \taujet
(\(p_{\tau}\)) or a light-jet (\(p_u\)) and is defined as:

\begin{equation*}
D_b = \log \left( \frac{p_b}{f_c p_c + f_{\tau} p_{\tau} + \left( 1 - f_c - f_{\tau}\right) p_u} \right).
\end{equation*}

A jet is considered $b$-tagged if it has a \(D_b\) score larger than a given
value. A selection on \(D_b\) defines an operating point (OP) associated with a
certain inclusive \bjet tagging efficiency, calculated as the fraction of
\bjets that are $b$-tagged. The mis-tagging rate for \(c\)-, \(\tau\)- and
\ljets is determined by the fraction of jets that are mistakenly $b$-tagged,
for that given jet flavour, and the rejection is the reciprocal of the
mis-tagging rate. The ATLAS Collaboration uses a sample of simulated
\(t\bar{t}\) events, where most jets have a \pT below \mbox{250 GeV}, to derive the
OPs. The free parameters \(f_{c (\tau)}\) determine the relative weighting
between \(p_{c (\tau)}\) and \(p_u\) in the discriminant \(D_b\). The specific
value of \(f_{c}\) is determined through an optimisation procedure aimed at
obtaining a certain balance between rejections of \cjets and \ljets\ in
simulated \ttbar events. The value of \(f_{\tau}\) is optimised to maximise the
\taujet rejection, while ensuring a negligible impact upon the \cjet and \ljet
rejection. In the case of GN2, \(f_{c (\tau)}\) is set to 0.2 (0.05), while for
DL1d, which does not have a \taujet output in the model, \(f_{c}\) is set to
0.018. For GN2, \(f_{c}\) is tuned to reach a much higher \cjet rejection,
while still achieving a better \ljet rejection, compared with DL1d.

\begin{figure}[!htb]
\centering
\begin{subfigure}{0.5\textwidth}
\includegraphics[width=\textwidth]{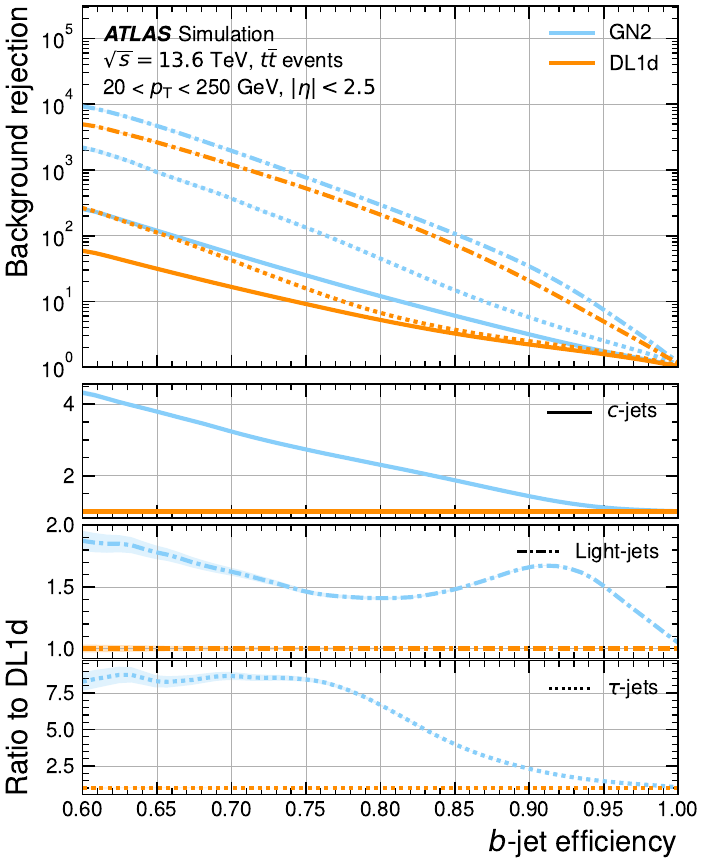}
\caption{}
\end{subfigure}%
\begin{subfigure}{0.5\textwidth}
\includegraphics[width=\textwidth]{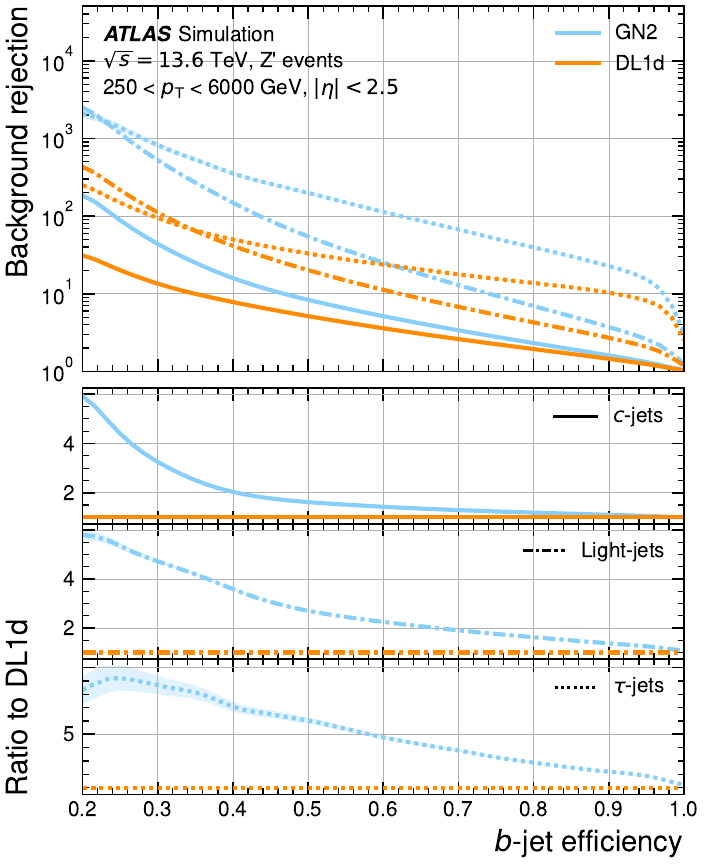}
\caption{}
\end{subfigure}
\caption{
The \cjet (solid), \ljet (dotted-dashed), and \taujet (dashed) rejections as a function of the \bjet tagging efficiency for (a) \ttbarjets with \ttbarpt and (b) \Zprimejets with \Zprimept, for both GN2 (light blue) and DL1d (dark orange).
The performance of GN2 with respect to DL1d is shown in the bottom panels.
The 68\% confidence intervals calculated assuming no correlations between the rejections are indicated by the shaded regions, and the uncertainty on each rejection is obtained according to a binomial distribution.
}
\label{fig:btag-rocs}
\end{figure}

Figure \ref{fig:btag-rocs} illustrates the tagger performance in terms of the
\cjet, \ujet and \taujet rejection as a function of the \bjet tagging
efficiency. In both the \ttbar and \Zprime samples, GN2 exhibits significantly
better background rejection compared with DL1d across the entire range of \bjet
tagging efficiencies. The degree of improvement depends on the \bjet tagging
efficiency. In the \ttbar sample, the \cjet (\ljet) rejection of GN2 improves
by more than a factor of 3 (1.6), compared with DL1d, for the most commonly used
\SI{70}{\percent} OP. The performance of both algorithms starts degrading once
the jet \pT reaches around 200 \GeV, due to several confounding factors, including
suboptimal tracking performance in dense environments where the spatial
separation between tracks becomes smaller~\cite{ctide}. In the \Zprime sample,
applying the \SI{70}{\percent} OP selection on \(D_b\) yields a \bjet tagging efficiency of
\SI{30}{\percent}, and the \cjet (\ljet) rejection of GN2 improves by more than
a factor of 3 (4), compared with DL1d. The inclusion of a \taujet output node in
GN2 leads to an even greater enhancement in the \taujet rejection, by up to a
factor of 8 (9) for jets in the \ttbar (\Zprime) sample, without
significantly degrading the \cjet and \ljet rejection.

The performance of a \(c\)-tagging algorithm is evaluated based on its ability
to reject \(b\)-, \(\tau\)- and \ljets while maintaining a desired \cjet
tagging efficiency. Due to the end-to-end architecture that does not rely on
low-level tagger inputs, GN2 can seamlessly be adapted as a $c$-tagging
algorithm without re-training any lower level algorithms. Similar to \btag, a
discriminant, $D_c$, is constructed as:

\begin{equation*}
D_c = \log \left( \frac{p_c}{f_b p_b + f_{\tau} p_{\tau} + \left( 1 - f_b - f_{\tau}\right) p_u} \right),
\end{equation*}

where $f_{b (\tau)}$ is the free parameter that controls the flavour
composition of the background in the background hypothesis. The value chosen for
$f_{b (\tau)}$ is 0.3 (0.01) for GN2, while for DL1d, \(f_{b}\) is 0.1,
following a similar optimisation procedure as for \(D_b\).

\begin{figure}[!htb]
\centering
\begin{subfigure}{0.5\textwidth}
\includegraphics[width=\textwidth]{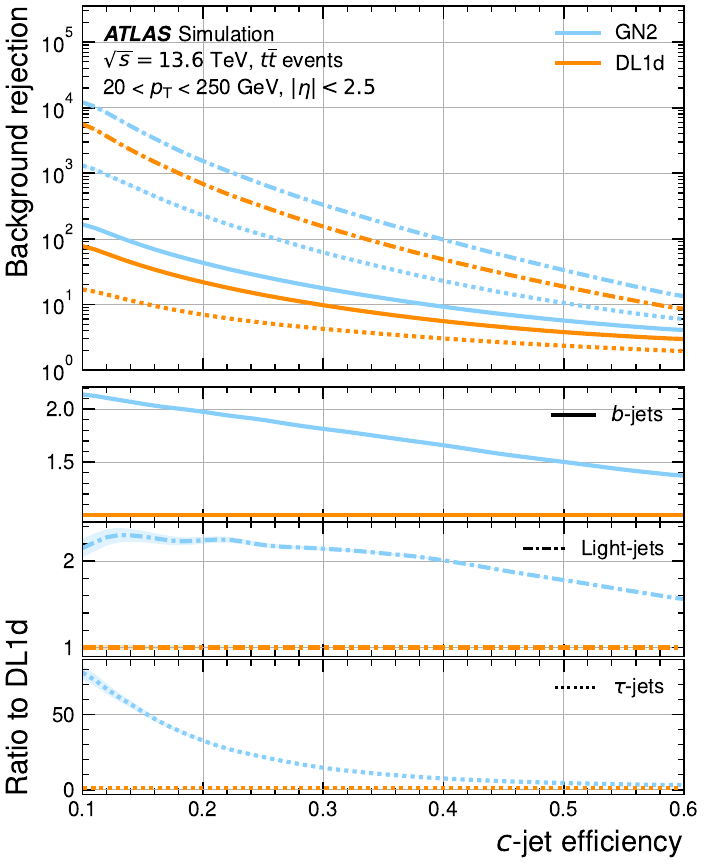}
\caption{}
\end{subfigure}%
\begin{subfigure}{0.5\textwidth}
\includegraphics[width=\textwidth]{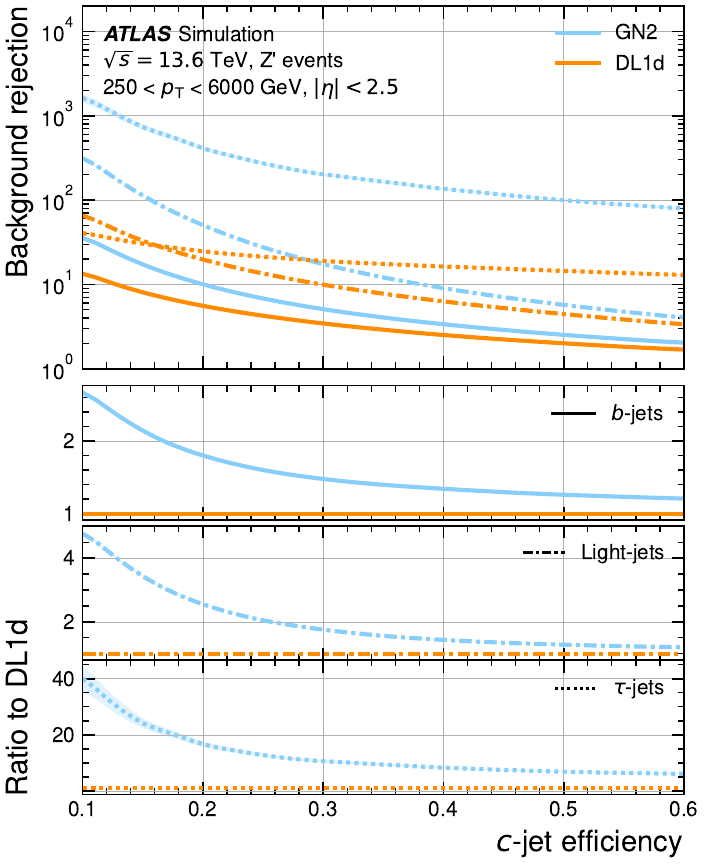}
\caption{}
\end{subfigure}
\caption{
The \bjet (solid), \ljet (dotted-dashed), and \taujet (dashed)  rejections as a function of the \cjet tagging efficiency for (a) \ttbarjets with \ttbarpt and (b) \Zprimejets with \Zprimept, for both GN2 (light blue) and DL1d (dark orange).
The performance of GN2 relative to DL1d is shown in the bottom panels.
The 68\% confidence intervals calculated assuming no correlations between the rejections are indicated by the shaded regions, and the uncertainty on each rejection is obtained according to a binomial distribution.
}
\label{fig:ctag-rocs}
\end{figure}

The \ctag performance of DL1d and GN2 are compared in \cref{fig:ctag-rocs},
which shows a significant improvement in performance across all \cjet tagging
efficiencies.  The \bjet (\ljet) rejection is enhanced by a factor of
approximately 1.8 (2.2) in the \ttbar sample at a \SI{30}{\percent} \cjet
tagging efficiency, which is a typical choice in measurements of the $c$-quark
Yukawa coupling~\cite{HIGG-2021-12}. The \bjet (\ljet) rejection is increased
by a factor of approximately 2.7 (4.7) in the \Zprime sample at a corresponding
efficiency of \SI{10}{\percent}. The \taurej is improved by a factor of
approximately 15 (40) in the \ttbar (\Zprime) sample.

\textbf{Algorithm performance in collision data}

\noindent Due to imperfections in the physics modelling of the MC
generator and in the simulated detector response, the distribution of the input variables to the
algorithms and their correlations differ between collision data and simulation,
resulting in a performance difference. It is not practical to correct each
individual mis-modelled variable, so dedicated calibration analyses are employed
to measure the tagging efficiency of \bjets, \cjets and light-jets for
pre-defined operating points (OPs)
directly~\cite{FTAG-2018-01,FTAG-2020-08,FTAG-2019-02}. In the case of the GN2
algorithm, five OPs are defined corresponding to inclusive \bjet tagging
efficiencies of 65$\%$, 70$\%$, 77$\%$, 85$\%$ and 90$\%$ while for DL1d four OPs
are constructed corresponding to inclusive \bjet tagging efficiencies of
60$\%$, 70$\%$, 77$\%$ and 85$\%$. The results presented in this paper are
derived using $pp$ collision data recorded during Run 2 of the LHC at $\sqrt s$
= 13 TeV, corresponding to an integrated luminosity of \mbox{140 $\mathrm{fb}^{-1}$}.
The tagging performance in data for \bjets, \cjets, and \ljets is measured, in
order to obtain jet-flavour-dependent simulation-to-data correction factors,
binned in jet \pT. They are applied to MC-simulated jets to rescale their
tagging efficiencies and mis-tagging rates to match those measured in collision
data. The calibration of \bjets and \cjets is done with $t\bar{t}$
events~\cite{FTAG-2018-01,FTAG-2020-08}, while the calibration of \ljets is
performed using jets produced in association with a \(Z\)
boson~\cite{FTAG-2019-02}. Details of the calibration analyses are provided in
the Methods section.

\begin{figure}[!t]
\centering
\begin{subfigure}{0.5\textwidth}
\includegraphics[width=\textwidth]{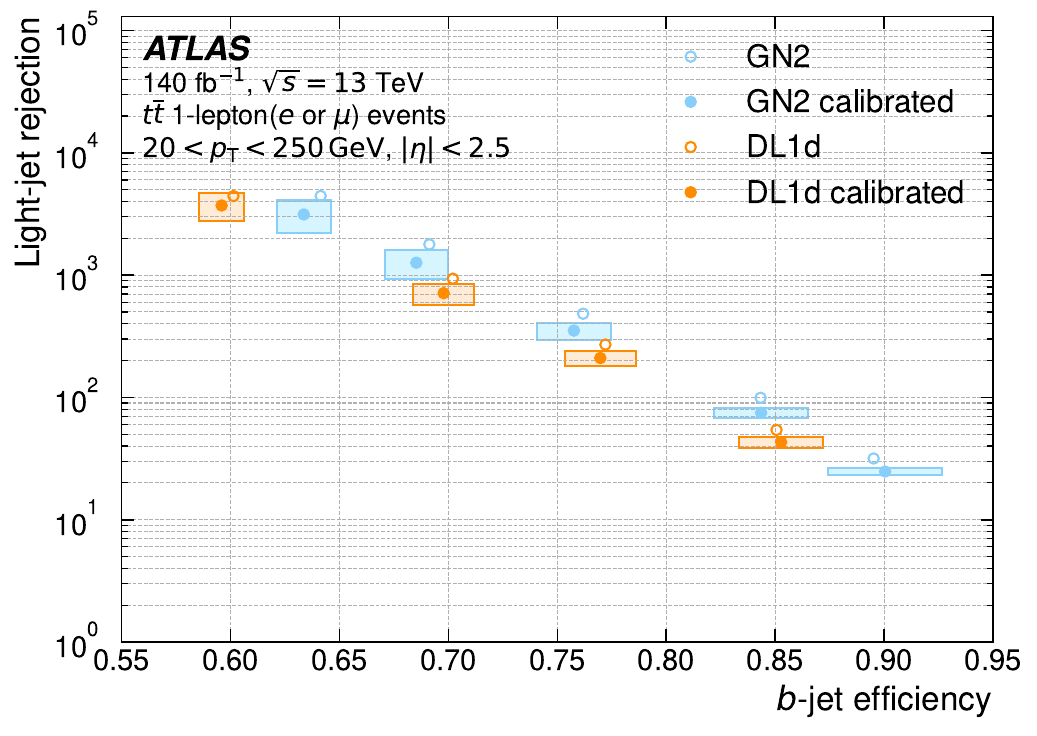}
\caption{}
\end{subfigure}%
\begin{subfigure}{0.5\textwidth}
\includegraphics[width=\textwidth]{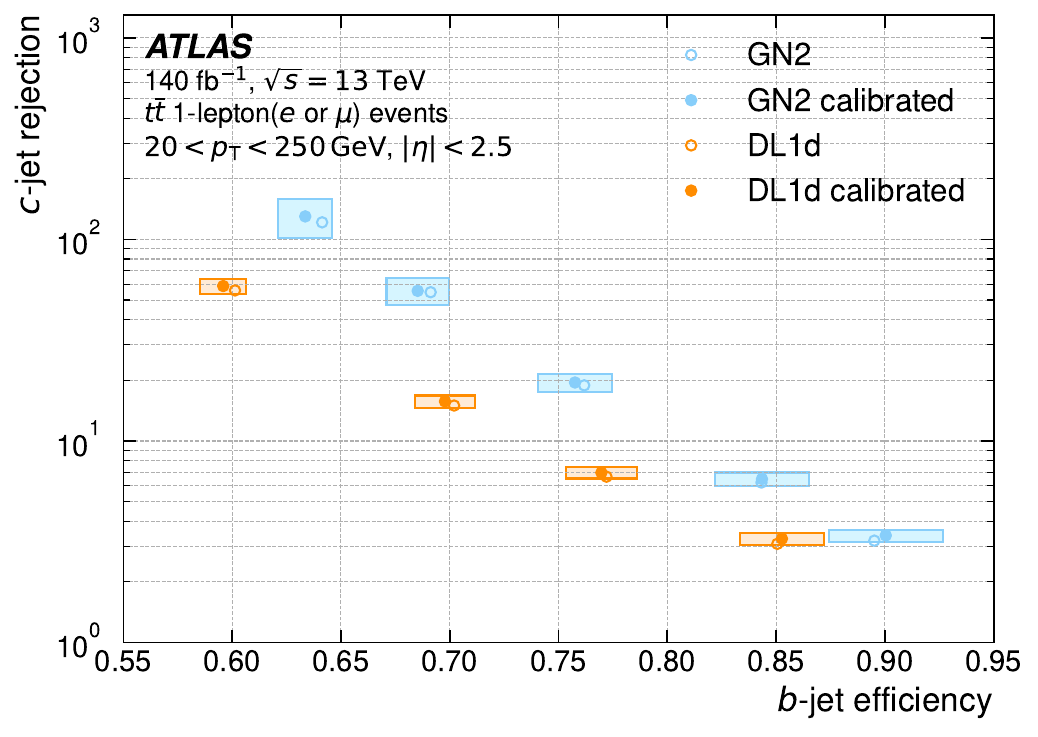}
\caption{}
\end{subfigure}
\caption{
The (a) \ljet rejection and (b) \cjet rejection as a function of the \bjet tagging efficiency for GN2 (light blue) and DL1d (dark orange), directly obtained in simulation (hollowed circle) and rescaled to match those in collision data  (solid point). The horizontal error bands correspond to the uncertainties associated with the \bjet tagging efficiency measurement, while the vertical error bands indicate the uncertainties associated with the rejection measurements. A \ttbar MC simulation sample with a reconstructed electron or muon is used to derive these results.
}
\label{fig:efficiency-rejection-calibrated}
\end{figure}

Figure~\ref{fig:efficiency-rejection-calibrated} presents the calibrated
tagging efficiencies and rejections of GN2 and DL1d, along with their
associated uncertainties, for each OP. The inclusive efficiencies and
rejections are obtained by averaging over the events in a simulated \ttbar
sample after requiring the presence of one reconstructed electron or muon. The
original efficiencies from the simulated sample are included as references,
enabling a direct comparison that shows similar agreement between data and
simulation for both GN2 and DL1d. The GN2 tagger demonstrates clear
improvements over DL1d in collision data. For instance, the measured \cjet (\ljet) rejection in data is increased by a factor of 3.5 (1.8) for the 70\% OP. The measurements in data provide conclusive evidence of the
enhanced performance enabled by advanced machine-learning algorithms in
identifying heavy-flavour jets at the LHC.

\FloatBarrier

% End of text imported from the .//sections/results.tex input file

%
%
%
%

%
%
\section*{Discussion}
\label{sec:discussion}
%

% The next lines are included from the .//sections/discussion.tex input file
Key challenges with machine-learning algorithms based on low-level inputs, such
as GN2, include the potential loss of interpretability and the need to ensure
consistent performance across different MC simulation methods. Robustness
against these potential shortcomings is critical to prevent the algorithm from
relying on unphysical features of the training sample.
In this section, these aspects are discussed further.

\textit{Physics inspiration and the auxiliary training objectives: }A key strength of the GN2 model lies in its physics-inspired constraints,
which aid the main task of jet classification while also improving the
interpretability of the model. This is accomplished by incorporating two
additional training objectives: predicting the origin of tracks associated with
the jet and determining which tracks originate from common vertices.
These objectives are not strictly necessary for the jet classification task and
are therefore referred to as auxiliary training objectives. The technical
implementation details are provided in the Methods section.

The track classification auxiliary training objective aims to estimate the
probability that a track originates from one of the following physical processes: a pile-up interaction~\cite{pileup}; the primary hard-scatter
interaction; the decay of a \bhadron; the decay of a \chadron produced by a
\bhadron; the decay of a \chadron; the decay of a \(\tau\)-lepton; or any
other secondary source. Class-weighted losses are applied during training to
mitigate the class imbalance, and tracks are classified by the
highest-probability category during evaluation. The class weights are fixed and based on the inverse class frequencies in the training dataset.
The classification efficiency refers to the probability for the track's origin
to be correctly predicted, in a group of tracks with certain target origins,
while the purity corresponds to the fraction of correctly predicted tracks,
within a group of tracks with specific predicted origins. When combining the
two categories involving a \bhadron, GN2 achieves an efficiency (purity) of
\SI{84}{\percent} (\SI{84}{\percent}). For tracks that are not of heavy-flavour (HF) origin, the efficiency (purity) is \SI{85}{\percent} (\SI{96}{\percent}). The above performance is evaluated in Run-3 samples at \(\sqs =\SI{13.6}{\tera\electronvolt}\).

The vertex finding auxiliary training objective aims to identify groups of
tracks that originate from a common spatial point. Each pair of tracks in the
jet is classified to determine whether they share the same vertex. Using these
pair-wise compatibility scores, track groups (vertices) are formed via a
union-find algorithm~\cite{Kozen1992}. SV1, an existing secondary vertex
reconstruction algorithm detailed in Ref.~\cite{FTAG-2019-07}, serves as a
reference algorithm. SV1 reconstructs a single inclusive vertex, whereas GN2
can identify multiple vertices of various types within a jet. Therefore, an
aggregation procedure is applied to the output of GN2 to enable a direct
comparison with the single inclusive vertex produced by SV1. To study the
vertex properties of \bjets, the identified vertex containing the most tracks
that have a predicted primary origin is removed, as this is likely to be the vertex associated with the primary hard-scatter interaction. Next, the remaining GN2 vertices that include at least one track predicted to have a HF
origin are consolidated into a single inclusive vertex.

\begin{figure}[!htb]
\centering
\begin{subfigure}{0.5\textwidth}
\includegraphics[width=\textwidth]{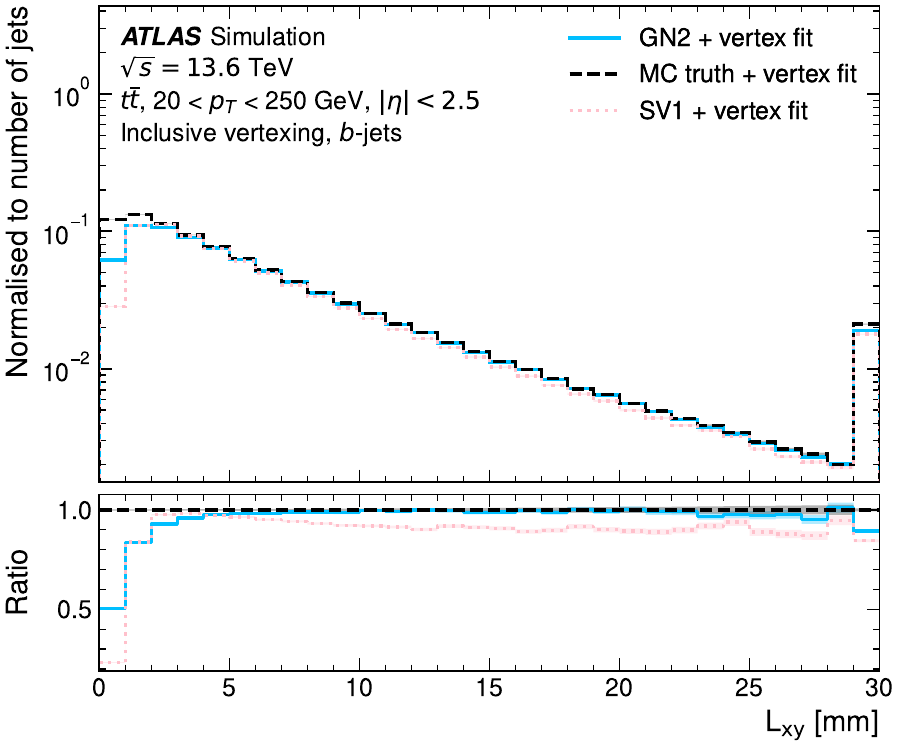}
\caption{}
\end{subfigure}%
\begin{subfigure}{0.5\textwidth}
\includegraphics[width=\textwidth]{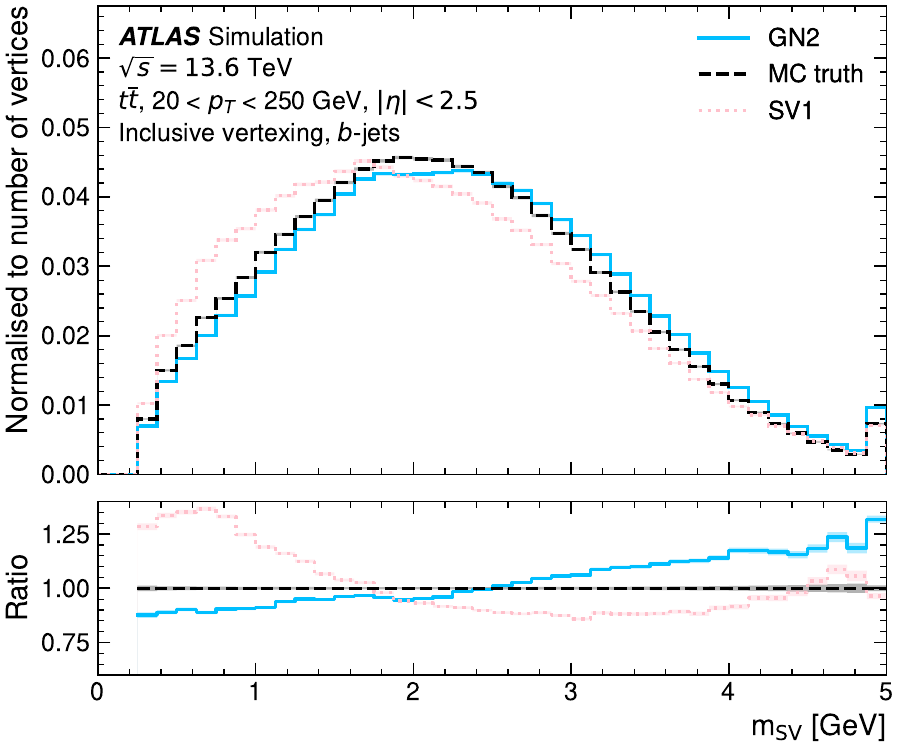}
\caption{}
\end{subfigure}
\caption{ The (a) transverse displacement and the (b) mass of the secondary vertex obtained by the GN2 (solid) and the SV1 (dotted) algorithms. While the transverse displacement is calculated via a Billoir fit performed on the tracks assigned to the vertex by the respective algorithm, the vertex mass is defined as the invariant mass of the same set of assigned tracks. MC truth (dashed) corresponds to an inclusive reference vertex derived from all tracks associated to simulation-level vertices containing only \bhadron tracks. The last bin in each plot includes overflow.
}
\label{fig:vert-perf}
\end{figure}

An inclusive reference vertex is constructed in simulated events, by combining
all tracks from simulation-level secondary vertices within the jet that consist
solely of HF tracks. A Billoir fit~\cite{Billoir} is performed on
the tracks selected by the GN2 and SV1 vertex finding algorithms to obtain the
transverse displacement of the vertex, $L_{xy}$. Figure~\ref{fig:vert-perf}
presents the $L_{xy}$ distribution for vertices obtained with GN2 and DL1d in
\bjets from a simulated \ttbar sample, compared to the expected distribution
derived from the inclusive reference vertex.  GN2 consistently achieves higher
vertex-finding efficiency than SV1 across the entire distribution of $L_{xy}$.
The mass of the secondary vertex can also be calculated using the momenta of
tracks selected by the vertex finding algorithms. The distribution of the
secondary vertex mass normalised to unity is also shown in
Figure~\ref{fig:vert-perf}. Remarkably, the mass of the secondary vertices
reconstructed by GN2 exhibits good agreement with the mass of the inclusive
reference vertex, despite the vertex mass not being explicitly targeted during
training. Unlike SV1, GN2 does not impose explicit selections on track
properties such as impact parameters. This leads to a higher efficiency, albeit
with a small contamination from non-HF tracks, which results in a slightly larger secondary vertex mass.

\begin{figure}[!htb]
\centering
\includegraphics[width=0.5\textwidth]{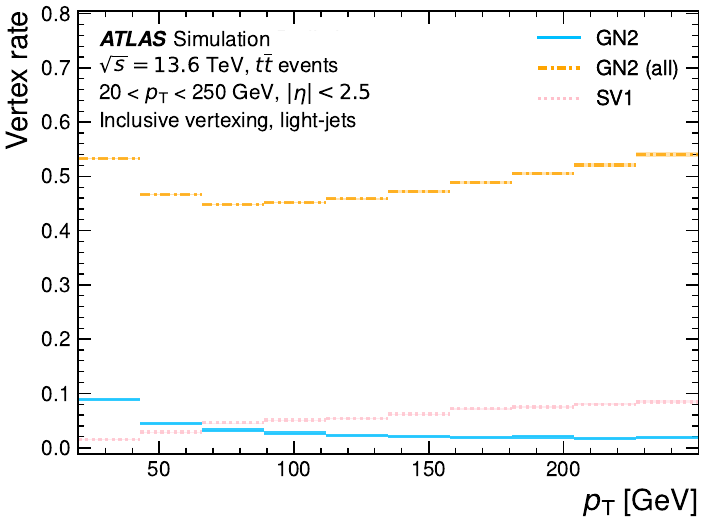}
\caption{
The rate of inclusive vertices reconstructed by the GN2 algorithm in \ljets as a function of the jet \pT, without any selections (dotted-dashed) and with the requirement of the vertex containing at least one track with predicted HF origin (solid). Results from the SV1 algorithm are added as a reference (dotted). The 68\% confidence intervals calculated according to a binomial distribution are indicated by the shaded regions.
}
\label{fig:light-rate}
\end{figure}

GN2 identifies all types of vertices including those from material
interactions, photon conversions, and in-flight decays of light hadrons.
Consequently, the rate of vertices reconstructed in \ljets, defined as the
fraction of \ljets containing a GN2 inclusive vertex, is expected to be much
higher compared to SV1 if no selections are applied in the aggregation
procedure described above. Figure~\ref{fig:light-rate} confirms this with
\ljets in the simulated \ttbar sample and shows that once requiring the GN2
inclusive vertex to contain at least one track with predicted HF
origin, the vertexing rate is dramatically reduced, down to the same level as
SV1.

\noindent To test the impact of the auxiliary objectives on the performance of
the main jet classification task, various GN2 configurations are
trained and tested. The resulting \cjet and \ljet rejections are reduced by up
to \SI{30}{\percent} in both the \ttbar and \Zprime samples, if both auxiliary
objectives are disabled. Disabling only one of them is sufficient to recover most of the performance loss, indicating that the two tasks are highly correlated in their contributions to the main jet flavour tagging objective.

Although the
outputs from the auxiliary tasks described above mainly serve as a way to
improve HF jet identification, with future development, their direct
usage in physics analyses remains a promising possibility.

\textit{Robustness against generator modelling variations: }Flavour-tagging algorithms are sensitive to the modelling of parton
showering, hadronisation, the underlying event and the properties of heavy-hadron decays~\cite{buckley}. To evaluate the robustness of the algorithm
against modelling variations, a comparative study of the GN2 performance in the
nominal simulated \ttbar sample used during training and samples produced with
alternative generator settings, both with Run-2 conditions at \(\sqs = \SI{13}{\tera\electronvolt}\), is performed.

The event and showering generators adopted for the nominal sample are
\POWHEGBOX~\cite{Frixione:2007nw,Nason:2004rx,Frixione:2007vw,Alioli:2010xd} and \PYTHIA~\cite{Sjostrand:2014zea}, respectively. The alternative samples include the use of a different showering
generator (\HERWIG~\cite{Herwig1,Herwig2,Herwig3}), whilst keeping the
same event generator, and the use of
\SHERPA~\cite{Bothmann:2019yzt} which applies a
different approach to all parts of the event generation model. The ratio
between the efficiency obtained with an alternative generator setup and with
the nominal setup is used to quantify the generator dependence of the
algorithms.

Table~\ref{tab:mcmc-mc23} shows these ratios for \bjets, \cjets, and \ljets at
the 70\% and 85\% OPs. Across the tested generators, the GN2 performance for
\bjets agrees to within 1--2\%, for \cjets the agreement is within 10\%, and
for light-jets the agreement is within 4\%. Similar agreement is also observed
for other OPs. The level of relative disagreement between DL1d and GN2 is close
to unity, suggesting that despite the GN2 model being significantly more
complex, it does not induce additional generator dependence.

%

%

% The next lines are included from the .//tables/mcmc-mc23.tex input file
\begin{table}[htbp]
\centering
\caption{Ratios of the efficiencies obtained with samples using alternative MC generators, relative to those in the nominal \POWHEGBOX + \PYTHIA sample used during training of the algorithm. Statistical uncertainties from evaluating the same algorithm on different samples are negligible and thus not shown.}
\label{tab:mcmc-mc23}
\setlength{\tabcolsep}{0.5em} %
\setlength{\extrarowheight}{2pt}
\begin{tabular}{ll cccc}
\toprule\hline
\multicolumn{2}{c}{} & \multicolumn{2}{c}{70\% OP} & \multicolumn{2}{c}{85\% OP} \\
\multicolumn{2}{c}{} & DL1d & GN2 & DL1d & GN2 \\
\hline
\multirow{3}{*}{\POWHEGBOX + \HERWIG} & \bjets & 0.984 & 0.984 & 0.989 & 0.990 \\
& \cjets & 0.951 & 0.904 & 0.977 & 0.983 \\
& \ujets & 1.003 & 1.000 & 1.015 & 1.011 \\
\hline
\multirow{3}{*}{\SHERPA}       & \bjets & 0.996 & 0.995 & 0.992 & 0.992 \\
& \cjets & 0.923 & 0.938 & 0.947 & 0.931 \\
& \ujets & 1.039 & 1.013 & 1.077 & 1.042 \\
\hline\bottomrule
\end{tabular}\vspace{1em}
\end{table}

% End of text imported from the .//tables/mcmc-mc23.tex input file

%

%
%
%
%
%
%
%
%
%
%
%
%

%
%
%

% End of text imported from the .//sections/discussion.tex input file

%
\section*{Methods}
\label{sec:methods}
%

% The next lines are included from the .//sections/methods.tex input file
\textbf{The ATLAS detector}

\newcommand{\AtlasCoordFootnote}{

\noindent ATLAS uses a right-handed coordinate system with its origin at the
nominal interaction point in the centre of the detector and the \(z\)-axis
along the beam pipe.  The \(x\)-axis points from the nominal interaction point to the centre of the
LHC ring, and the \(y\)-axis points upwards.  Polar coordinates \((r,\phi)\)
are used in the transverse plane, \(\phi\) being the azimuthal angle around the
\(z\)-axis.  The pseudorapidity is defined in terms of the polar angle
\(\theta\) as \(\eta = -\ln \tan(\theta/2)\) and is equal to the rapidity $ y =
\frac{1}{2} \ln \left( \frac{E + p_z c}{E - p_z c} \right) $ in the
relativistic limit.  Angular distance is measured in units of \(\Delta R \equiv
\sqrt{(\Delta \eta)^{2} + (\Delta\phi)^{2}}\).}

The ATLAS experiment~\cite{PERF-2007-01,GENR-2019-02} at the LHC is a
multipurpose particle detector with a forward–backward symmetric cylindrical
geometry and a solid-angle coverage of almost \(4\pi\). It is used to record
particles produced in $pp$ collisions at the LHC through a combination of particle position
and energy measurements. It includes an inner-tracking detector (ID)
surrounded by a thin superconducting solenoid providing a \SI{2}{\tesla} axial
magnetic field, electromagnetic and hadronic calorimeters, and a muon
spectrometer. The ID consists of silicon pixel, silicon microstrip, and
transition radiation tracking detectors. The muon spectrometer surrounds the
calorimeters and is based on three large superconducting air-core toroidal
magnets with eight coils each providing a field integral of between
\SI{2}{\tesla\meter} and \SI{6}{\tesla\meter} across the detector.

An extensive software suite~\cite{SOFT-2022-02} is used in data simulation, the
reconstruction and analysis of real and simulated data, detector operations,
and the trigger and data acquisition systems of the experiment.

\textbf{Monte Carlo simulation samples}

The \ttbar events at \(\sqs = \SI{13}{\tera\electronvolt}\) are modelled using the
\POWHEGBOX[v2]~\cite{Frixione:2007nw,Nason:2004rx,Frixione:2007vw,Alioli:2010xd}
event generator at next-to-leading-order (NLO) in the strong coupling constant
\alphas with the \NNPDF[3.0nlo]~\cite{Ball:2014uwa} parton distribution
function (PDF) set and the first-gluon-emission cut-off scale parameter \hdamp
set to \(1.5m_t\), with a top-quark mass of \(m_t = \SI{172.5}{\GeV}\). Parton
shower, hadronisation, and the underlying event are modelled by interfacing
\POWHEGBOX[v2] to \PYTHIA[8.230]~\cite{Sjostrand:2014zea},
using the A14 set of tuned parameters~\cite{ATL-PHYS-PUB-2014-021} and the
\NNPDF[2.3lo] PDF set~\cite{Ball:2012cx}. The decays of $b$- and $c$-hadrons are performed by \EVTGEN[1.6.0]~\cite{Lange:2001uf}.

The \Zprime events at \(\sqs = \SI{13}{\tera\electronvolt}\)
used to enrich the dataset with high-\pt jets are generated using
\Pythia[8.243] with the A14 set of tuned parameters for the underlying event
and the leading-order (LO) \NNPDF[2.3lo] PDF set. A broad jet \pt spectrum with
an almost uniform distribution between 250~\GeV\ and 1.5~\TeV\ and a tail
expanding to 6~\TeV\ is obtained by applying a weighting factor that modifies the original cross-section of the \Zprime resonance. The decays to \(b\bar b\), \(c \bar
c\), and light-flavour quark pairs are set to have equal branching ratios,
while the branching ratio to \(\tau\bar \tau\) is set to 5\%. The decays of
$b$- and $c$-hadrons are performed by \EVTGEN[1.7.0].

The \ttbar and \Zprime events at \(\sqs = \SI{13.6}{\tera\electronvolt}\) are
produced using the same setups, but with newer versions of \PYTHIA (8.308) and
\EVTGEN (2.1.1).

The impact of using different generators and models for parton shower
and hadronisation is studied with simulated \ttbar events from alternative
generator setups. Two scenarios are considered, where either only the showering
algorithm is varied, or the entire chain is changed. The former is achieved by
interfacing the \POWHEGBOX[v2] generator with the \HERWIG[7.2.1]~\cite{Herwig1,Herwig2,Herwig3} showering algorithm
using the \HERWIG[7.1] default set of tuned parameters, with the \NNPDF[3.0nlo]
set of PDFs. The latter is realised with the
\SHERPA[2.2.12]~\cite{Bothmann:2019yzt} generator, using NLO-accurate matrix
elements for up to one additional parton, and LO-accurate matrix elements for
up to four additional partons, calculated with the
\textsc{COMIX}~\cite{Gleisberg:2008fv} and
\OPENLOOPS~\cite{Buccioni:2019sur,Cascioli:2011va,Denner:2016kdg} libraries.
The \SHERPA parton shower~\cite{Gleisberg_2009,Schumann:2007mg} is applied using the \MEPSatNLO
prescription~\cite{Hoeche:2011fd,Hoeche:2012yf,Catani:2001cc,Hoeche:2009rj} and
the set of tuned parameters developed by the \SHERPA authors to match the
\NNPDF[3.0nnlo] set of PDFs.

\textbf{Objects for flavour tagging}

ATLAS uses a right-handed coordinate system with its
origin at the nominal interaction point in the centre of the detector and the \(z\)-axis along the beam pipe.
The \(x\)-axis points from the nominal interaction point to the centre of the LHC ring, and the \(y\)-axis
points upwards.
Cylindrical coordinates \((r,\phi)\) are used in the transverse plane, $\phi$
being the azimuthal angle around the \(z\)-axis.
The pseudorapidity is defined in terms of the polar angle $\theta$ as
\(\eta = -\ln \tan(\theta/2)\). Angular distance is measured in units of
\(\DeltaR \equiv \sqrt{(\Delta\eta)^{2} + (\Delta\phi)^{2}}\).

\noindent The fundamental objects for flavour tagging are jets, tracks, and vertices.
A concise description of these objects is provided
below, while a detailed description is available in Ref.~\cite{FTAG-2019-07}.

Tracks are reconstructed from ID information~\cite{ctide,IDTR-2022-04}.
To be considered for jet flavour tagging they are required to be within
\(|\eta| < 2.5\), have \(\pt > \SI{0.5}{\giga\electronvolt}\) and satisfy
criteria designed to reject fake and poorly measured
tracks~\cite{ATL-PHYS-PUB-2022-033}.

Primary vertices (PVs) are reconstructed from tracks in the luminous region of
the colliding LHC beams using an adaptive multi-vertex
filter~\cite{PERF-2015-01,ATL-PHYS-PUB-2015-026}. The PV with the highest sum
of squared transverse momenta \pt of contributing tracks is selected as the
primary interaction point (IP) and provides the reference point in an event.
The distance of closest approach of a track to the IP, the \enquote{perigee},
is indicated in the transverse plane by the transverse impact parameter
\(d_{0}\). The longitudinal separation between the IP and the point on the
track where \(d_0\) is measured, is indicated by the longitudinal impact
parameter \(z_0\). Tracks with large impact parameters can indicate the
presence of displaced decays, providing vital information to the flavour
tagging algorithms.

The DIPS and GN2 algorithms require tracks to be reconstructed from at least 8
hits in the silicon detector, at most one of which contributes to two tracks,
at most two \enquote{holes} in the silicon detector, and at most one hole in
the pixel detector, where hole denotes a hit missing where one is expected from
the track trajectory.  Further, requirements on the track impact parameters, \(|d_0| < \SI{3.5}{\milli\meter}\) and \(|z_0 \sin \theta| <
\SI{5}{\milli\meter}\), retain charged particle tracks originating from
HF hadron decays while suppressing tracks from other sources.

Jets are reconstructed using the anti-\kt algorithm \cite{Cacciari:2008gp} with
radius parameter \(R=0.4\) using the `fastjet' package \cite{Fastjet}. The
input constituents are \enquote{particle-flow} objects~\cite{PERF-2015-09}
which combine signals in the ATLAS calorimeters and ID to exploit precision
tracking information for low-\pt charged hadrons spatially matched with
calorimeter energy deposits. The jet \pt is corrected to the corresponding
particle-level jet \pt using calibration techniques described in
Ref.~\cite{JETM-2018-05}.  The jets are required to have \(\pt >
\SI{20}{\giga\electronvolt}\) (to be within the valid calibration range) and
\(|\eta| < 2.5\) (to be within the tracking fiducial volume set by the ID
acceptance) to be considered for flavour tagging. Additionally, jets from
pile-up interactions are suppressed by the \enquote{jet vertex tagger} (JVT)
algorithm~\cite{ATLAS-CONF-2014-018}, which uses the ID tracks associated with
the jet to form a multivariate discriminant.  The JVT efficiency for jets
originating from the IP is \SI{92}{\percent} in the simulation.  The jet axis,
derived from the sum of the momenta of the jet constituents, is used when
associating tracks with the jet and when assigning a lifetime sign to the
tracks' impact parameters. Tracks are associated with a given jet by setting a
maximum allowed angular separation \DeltaR between the track momenta, defined
at the perigee, and the jet axis. The \DeltaR requirement varies as a function
of the jet \pt to account for decay products from \(b\)-hadrons with larger \pt
being more collimated, ranging from \(0.45\) for jet
\(\pt=\SI{20}{\giga\electronvolt}\) to \(0.26\) for jet \(\pt >
\SI{150}{\giga\electronvolt}\). If a track can be associated with multiple
jets, it is assigned to the jet closest in \DeltaR. The sign convention for the
lifetime-signed impact parameters assigns a positive sign if the track
intersects the jet axis in the transverse plane in front of the IP, and a
negative sign if the intersection lies behind the IP~\cite{FTAGRun1}.  The
flavour labels of jets in simulation are assigned depending on the hadrons
associated with the jet. The set of weakly decaying hadrons and hadronically
decaying \tauleps with \(\pt > \SI{5}{\giga\electronvolt}\) within a \(\DeltaR
< 0.3\) cone around the jet axis determines the jet flavour following a
sequential labelling decision tree.  A jet is labelled a \bjet if it contains
at least one \bhadron with \(\pt > \SI{5}{\giga\electronvolt}\), a \cjet
(\taujet) if it contains at least one \chadron (hadronic \taulep decay) and no \bhadron, and
otherwise it is called a \ljet, where the latter is an inclusive label for the
jets originating from a light quark or gluon. These labels are used both for
training the algorithms, and for evaluating their performance.

Targets for the auxiliary training objectives are obtained from the
simulation-level event record. Tracks are matched with simulation-level
particles using the approach in Ref.~\cite{ctide}. Track-origin labels are
obtained by analysing the decay history of the matched particles, while
track-pair-compatibility labels are obtained by considering the production
vertices of the matched particles. Production vertices within $\SI{0.1}{\mm}$
in 3D space are merged to account for the finite resolution of the detector,
and the matched track-pairs are assigned the same label.

\textbf{The algorithm architecture}

\noindent The primary flavour tagging algorithm presented is GN2, which
directly learns from the charged particle
tracks via a transformer-based model. Another algorithm, DL1d, which follows
previous approaches of combining inputs from several low-level taggers in a
multivariate technique, is also discussed as a baseline reference.

Both algorithms are trained on a dataset created from combining the simulated
\ttbar and \Zprime samples described earlier in this section. Jets with
\(\SI{20}{\giga\electronvolt} < \pt < \SI{250}{\giga\electronvolt}\) are taken
from the \ttbar sample and those with \(\SI{250}{\giga\electronvolt} < \pt <
\SI{6}{\tera\electronvolt}\) from the \Zprime sample. The \bjets, \ljets and
\taujets are re-sampled in \pt and \(\eta\) to match the corresponding \cjet
distributions, thereby preventing the models from discriminating between jet
flavours based on relative kinematic differences.  All input variables to the
algorithm training are normalised to have zero mean and unit variance. A coarse
optimisation of hyperparameters, such as the number of layers, is carried out for both algorithms, and the
AdamW \cite{AdamW} (Adam~\cite{kingma2014adam}) optimizer is used for
training GN2 (DL1d) with the learning rate and optimisation schedule defined
below.

The GN2 algorithm is an end-to-end architecture without any intermediate
taggers involved, as illustrated in Figure~\ref{fig:gn2_sketch}. It is based on
the GN1~\cite{ATL-PHYS-PUB-2022-027} demonstrator version of the algorithm,
replacing the Graph Attention Network~\cite{gatv2} with a
Transformer~\cite{transformers} along with other architecture optimisations.
GN2 directly accepts information about the jet and associated tracks that are
provided by the standard event reconstruction. This results in a simpler and
more flexible algorithm which can be easily reoptimised for different physics
objectives, such as the identification of highly energetic Higgs bosons decaying into $b$- or $c$-quark pairs~\cite{ATL-PHYS-PUB-2023-021}, jet energy regression~\cite{ATL-PHYS-PUB-2024-015}, exotic jet tagging~\cite{EMJ3}, and jet flavour tagging in the ATLAS high-level trigger~\cite{TRIG-2022-02}. Additionally, when compared with DL1d, GN2 is
trained to recognise an additional class of jets that originate from hadronic
\(\taulep\) decays.

First, the jet features are concatenated with a fixed-size array of 40 track feature vectors, with unused elements masked when fewer than 40 tracks are available, allowing it to handle variable track multiplicity without zero-padding. Tracks with smaller absolute track impact parameter significance~\cite{FTAG-2019-07} are dropped if there are more than 40 tracks. The same inputs as for
GN1~\cite{ATL-PHYS-PUB-2022-027} are used, except the variables related to
holes in the silicon tracker, which were found to have no impact on
performance. A complementary interpretability analysis using integrated
gradients shows that the impact parameter significances and angular variables
emerge as particularly influential~\cite{ATL-PHYS-PUB-2025-029}. The combined
vectors are then fed into a per-track initialisation network, which is composed
of a single hidden layer and an output layer of size 256. Next, a four layer
transformer encoder with eight attention heads is used to produce track
representations that incorporate information from other tracks inside the jet.
The transformer has an embedding size of 256 and a feed-forward dimension of
512, and uses pre-LayerNorm~\cite{pre-ln}. After the transformer encoder, the
output track representations are projected down to dimension 128, and a global
representation of the jet is produced using attention
pooling~\cite{global-attention-pooling}. The pooled jet representation and
output track embeddings are provided as inputs to the three task-specific
networks. The primary objective, jet classification, uses only the pooled jet
representation and has an output layer of size 4, providing \(p_b\), \(p_c\),
\(p_u\) and \(p_\tau\) for the final discriminant definition.  The two
auxiliary objectives introduced in the Discussion section take advantage of the
track embeddings, in addition to the global jet representation. The track
origin classification task uses individual track embeddings and has 7 output
categories, while the track-pair compatibility task employs a binary output
layer, using the embeddings of each pair of tracks. Each task-specific network
consists of three hidden layers with size 128, 64 and 32, respectively. ReLU
activation~\cite{relu} is used throughout the model. Cross-entropy loss is used
by all three task-specific networks, which is combined with tunable weights to
form the final loss function, enabling a simultaneous optimisation of the
entire algorithm. GN2 applies the same auxiliary network structures and loss weights as GN1~\cite{ATL-PHYS-PUB-2022-027}.

GN2 is trained using a 4-fold strategy to prevent memorisation of the training
samples, given their possible use in ATLAS physics analyses.  Jets are assigned
to one of the four folds pseudo-randomly, with a number seeded by the event
number and discrete jet properties.  Four classifiers are then trained, each
excluding one of the four folds from the training dataset.  In physics
analysis, each jet is tagged using the classifier it was excluded from during
training. Each of the four networks has approximately 2.3M trainable parameters
and is trained using approximately 45M (18M) $b$-jets, 45M (18M) $c$-jets, 90M
(36M) light-jets and 6.25M (2.5M) $\tau$-jets from the \ttbar (\Zprime) sample,
simulated at both \(\sqs = \SI{13}{\tera\electronvolt}\) and \(\sqs =
\SI{13.6}{\tera\electronvolt}\), with a mixing ratio of 2:1. A learning rate
scheduler with cosine annealing~\cite{Loshchilov:2016tgk} is used with the
initial learning rate set to $1\times10^{-7}$, which is increased to
$5\times10^{-4}$ after the first \pct{1} of training steps have been completed.
It reduces to $1\times10^{-5}$ over the remainder of the training run. A weight
decay of $1\times10^{-5}$ is also added. A batch size of 12000 is adopted. The
different folds have compatible performance within statistical uncertainty. The
training data is translated from a standard ATLAS format~\cite{SOFT-2022-02} to
\textsc{HDF5}~\cite{hdf5}. The network is trained with \textsc{PyTorch
Lightning}~\cite{pytorch,lightning,salt2025}, consuming roughly 300 GPU hours on an
NVIDIA A100 card. It is deployed in ATLAS software with
\textsc{OnnxRuntime}~\cite{onnxruntime}, adding negligible CPU time. With the updated architecture and training setup, the \cjet(\ljet) rejection is improved by a factor of 1.5 (1.7) for a 70\% \bjet tagging efficiency, in the \ttbar sample, and by a factor of 1.3 (1.4) for the corresponding 30\% \bjet tagging efficiency, in the \Zprime sample.

The DL1d algorithm inherits the architecture from its predecessor
DL1r, described in Ref.~\cite{FTAG-2019-07}, but processes track impact
parameters with the DIPS algorithm based on
DeepSets~\cite{Komiske_2019,ATL-PHYS-PUB-2020-014} instead of a recurrent
neural network~\cite{ATL-PHYS-PUB-2017-003}. Overall, 44 input features are fed
into DL1d, including the jet \pt and \(\eta\). The architecture of DL1d
includes eight hidden layers of size 256, 128, 60, 48, 36, 24, 12, and 6, each
followed with ReLU activation and batch normalisation. The training was
performed with a learning rate of \num{1e-3} and training batch size of 15,000.
The training data pipeline is similar to GN2, with the exception that training
is done with \textsc{Keras} and \textsc{Tensorflow}~\cite{tensorflow} via
\textsc{Umami}, a dedicated Python toolkit~\cite{Umami}, and deployed in the
ATLAS software with \textsc{lwtnn}~\cite{lwtnn}.

\textbf{Performance measurement strategies in collision data}

The measurement of the \bjet tagging efficiency in collision data is carried
out in a roughly 90\% pure sample of \(t\bar{t}\) events where both top quarks
decay leptonically into a lepton, a neutrino and a $b$-quark. The events are
required to contain exactly one electron and one muon of opposite charge, in
addition to two jets. The invariant masses of the two lepton-jet pairs are used
to define one region enriched in \bjets and three control regions (CRs). The
\bjet-enriched region is determined by requiring that both lepton-jet pairs
have invariant masses compatible with an on-shell top quark decay. The CRs are
used in a likelihood fit to constrain the predicted jet flavour composition.
They are constructed to have increased fractions of non-\bjets by requiring
that at least one or both of the lepton-jet pairs do not originate from the
same top-quark decay. The analysis employs a statistical model based on a
likelihood function that extracts the efficiency in collision data binned in
\pT for all the \bjets in the sample. The dominant systematic uncertainty comes
from the modelling of \(t\bar{t}\) events. Additional details on the \bjet
calibration procedure are available in Ref.~\cite{FTAG-2018-01}.

The calibration measurement of the \cjet mis-tagging rate is performed in
$t\bar{t}$ events where one top quark decays leptonically while the other top
quark decays hadronically. A sample of \cjets is obtained through the
$W^{\pm}\rightarrow c\bar{s} (\bar{c}s)$ decay from the hadronically decaying
top quark. A likelihood-based kinematic reconstruction is employed to find,
among the four jets in the event, two jets associated with the hadronically
decaying \(W\)-boson and two jets stemming from the $b$-quarks produced in the
top quark decays.  The mis-tagging rate of \(c\)-jets is determined by
minimizing a \(\chi^2\) function computed in bins of the jet \pT of the two
jets from the \(W\)-boson decay. Additional terms that correct for
the potential mis-modelling of the total number of events in each jet \pT bin are estimated
simultaneously from the fit to collision data, while the contribution of background
events, in which no \cjets are associated with the \(W\)-boson decay, is
estimated from simulations. The mis-tagging rate of \ljets in this sample is
corrected using the method described below. As with the \bjets calibration analysis, the leading source of systematic uncertainties is the modelling of \(t\bar{t}\) events. The \cjet
mis-tagging rate calibration procedure is detailed in Ref.~\cite{FTAG-2020-08}.

The mis-tagging rate for \ljets is determined using jets produced in
association with a \(Z\) boson, where the \(Z\) boson decays into muon or
electron pairs. The key challenge in this calibration is to develop a method
capable of extracting a light-jet mis-tagging rate in data despite the high
rejection of the taggers. The method used in this work involves exploiting
transformed track variables in alternate taggers that provide reduced
$b$($c$)-jet tagging efficiency and almost unchanged \ljet rejection. The
mis-tagging rate of this modified tagger is measured from a fit to the
flavour-sensitive secondary vertex mass distribution in collision data, and
dedicated uncertainties are introduced so that it can be extrapolated to that
of the nominal tagger. These extrapolation uncertainties are a leading source of systematic uncertainty. A detailed description of the procedure is provided in
Ref.~\cite{FTAG-2019-02}.

\textbf{Data availability}

Raw data were generated by the ATLAS experiment. Derived data supporting the
findings of this study are available from the ATLAS Collaboration upon request.
A subset of the training sample and instructions to train GN2 can be
acquired via the CERN Open Data
Portal~\cite{OpenData,jetset-examples}.

\textbf{Code availability}

The ATLAS data reduction software is available at \textit{Zenodo}
(\url{https://doi.org/10.5281/zenodo.4772550})~\cite{athena}. The GN2 training
software suite can be found at \textit{JOSS}
(\url{https://joss.theoj.org/papers/10.21105/joss.07217})~\cite{salt2025}, and
the DL1d training software stack is available at \textit{JOSS}
(\url{https://joss.theoj.org/papers/10.21105/joss.05833})~\cite{Umami}.

% End of text imported from the .//sections/methods.tex input file

%
\section*{Acknowledgements}
%

% The next lines are included from the .//acknowledgements/Acknowledgements.tex input file
%
%

%
%

We thank CERN for the very successful operation of the LHC and its injectors, as well as the support staff at
CERN and at our institutions worldwide without whom ATLAS could not be operated efficiently.

The crucial computing support from all WLCG partners is acknowledged gratefully, in particular from CERN, the ATLAS Tier-1 facilities at TRIUMF/SFU (Canada), NDGF (Denmark, Norway, Sweden), CC-IN2P3 (France), KIT/GridKA (Germany), INFN-CNAF (Italy), NL-T1 (Netherlands), PIC (Spain), RAL (UK) and BNL (USA), the Tier-2 facilities worldwide and large non-WLCG resource providers. Major contributors of computing resources are listed in Ref.~\cite{ATL-SOFT-PUB-2025-001}.

We gratefully acknowledge the support of ANPCyT, Argentina; YerPhI, Armenia; ARC, Australia; BMWFW and FWF, Austria; ANAS, Azerbaijan; CNPq and FAPESP, Brazil; NSERC, NRC and CFI, Canada; CERN; ANID, Chile; CAS, MOST and NSFC, China; Minciencias, Colombia; MEYS CR, Czech Republic; DNRF and DNSRC, Denmark; IN2P3-CNRS and CEA-DRF/IRFU, France; SRNSFG, Georgia; BMFTR, HGF and MPG, Germany; GSRI, Greece; RGC and Hong Kong SAR, China; ICHEP and Academy of Sciences and Humanities, Israel; INFN, Italy; MEXT and JSPS, Japan; CNRST, Morocco; NWO, Netherlands; RCN, Norway; MNiSW, Poland; FCT, Portugal; MNE/IFA, Romania; MSTDI, Serbia; MSSR, Slovakia; ARIS and MVZI, Slovenia; DSI/NRF, South Africa; MICIU/AEI, Spain; SRC and Wallenberg Foundation, Sweden; SERI, SNSF and Cantons of Bern and Geneva, Switzerland; NSTC, Taipei; TENMAK, T\"urkiye; STFC/UKRI, United Kingdom; DOE and NSF, United States of America.

Individual groups and members have received support from BCKDF, CANARIE, CRC and DRAC, Canada; CERN-CZ, FORTE and PRIMUS, Czech Republic; COST, ERC, ERDF, Horizon 2020, ICSC-NextGenerationEU and Marie Sk{\l}odowska-Curie Actions, European Union; Investissements d'Avenir Labex, Investissements d'Avenir Idex and ANR, France; DFG and AvH Foundation, Germany; Herakleitos, Thales and Aristeia programmes co-financed by EU-ESF and the Greek NSRF, Greece; BSF-NSF and MINERVA, Israel; NCN and NAWA, Poland; La Caixa Banking Foundation, CERCA Programme Generalitat de Catalunya and PROMETEO and GenT Programmes Generalitat Valenciana, Spain; G\"{o}ran Gustafssons Stiftelse, Sweden; The Royal Society and Leverhulme Trust, United Kingdom.

In addition, individual members wish to acknowledge support from Armenia: Yerevan Physics Institute (FAPERJ); CERN: European Organization for Nuclear Research (CERN DOCT); Chile: Agencia Nacional de Investigaci\'on y Desarrollo (FONDECYT 1230812, FONDECYT 1240864); China: Chinese Ministry of Science and Technology (MOST-2023YFA1605700, MOST-2023YFA1609300), National Natural Science Foundation of China (NSFC - 12175119, NSFC 12275265); Czech Republic: Czech Science Foundation (GACR - 24-11373S), Ministry of Education Youth and Sports (ERC-CZ-LL2327, FORTE CZ.02.01.01/00/22\_008/0004632), PRIMUS Research Programme (PRIMUS/21/SCI/017); EU: H2020 European Research Council (ERC - 101002463); European Union: European Research Council (BARD No. 101116429, ERC - 948254, ERC 101089007), European Regional Development Fund (SMASH COFUND 101081355, SLO ERDF), Horizon 2020 Framework Programme (MUCCA - CHIST-ERA-19-XAI-00), European Union, Future Artificial Intelligence Research (FAIR-NextGenerationEU PE00000013), Italian Center for High Performance Computing, Big Data and Quantum Computing (ICSC, NextGenerationEU); France: Agence Nationale de la Recherche (ANR-21-CE31-0022, ANR-22-EDIR-0002); Germany: Baden-Württemberg Stiftung (BW Stiftung-Postdoc Eliteprogramme), Deutsche Forschungsgemeinschaft (DFG - 469666862, DFG - CR 312/5-2); China: Research Grants Council (GRF); Italy: Istituto Nazionale di Fisica Nucleare (ICSC, NextGenerationEU), Ministero dell'Università e della Ricerca (NextGenEU 153D23001490006 M4C2.1.1, NextGenEU I53D23000820006 M4C2.1.1, NextGenEU I53D23001490006 M4C2.1.1, SOE2024\_0000023); Japan: Japan Society for the Promotion of Science (JSPS KAKENHI JP22H01227, JSPS KAKENHI JP22H04944, JSPS KAKENHI JP22KK0227, JSPS KAKENHI JP23KK0245, JSPS KAKENHI JP24K23939); Norway: Research Council of Norway (RCN-314472); Poland: Ministry of Science and Higher Education (IDUB AGH, POB8, D4 no 9722), Polish National Science Centre (NCN 2021/42/E/ST2/00350, NCN OPUS 2023/51/B/ST2/02507, NCN OPUS nr 2022/47/B/ST2/03059, NCN UMO-2019/34/E/ST2/00393, UMO-2022/47/O/ST2/00148, UMO-2023/49/B/ST2/04085, UMO-2023/51/B/ST2/00920, UMO-2024/53/N/ST2/00869); Portugal: Foundation for Science and Technology (FCT); Spain: Ministry of Science and Innovation (MCIN \& NextGenEU PCI2022-135018-2, MICIN \& FEDER PID2021-125273NB, RYC2019-028510-I, RYC2020-030254-I, RYC2021-031273-I, RYC2022-038164-I); Sweden: Carl Trygger Foundation (Carl Trygger Foundation CTS 22:2312), Swedish Research Council (Swedish Research Council 2023-04654, VR 2021-03651, VR 2022-03845, VR 2022-04683, VR 2023-03403, VR 2024-05451), Knut and Alice Wallenberg Foundation (KAW 2018.0458, KAW 2022.0358, KAW 2023.0366); Switzerland: Swiss National Science Foundation (SNSF - PCEFP2\_194658); United Kingdom: Leverhulme Trust (Leverhulme Trust RPG-2020-004), Royal Society (NIF-R1-231091); United States of America: U.S. Department of Energy (ECA DE-AC02-76SF00515), Neubauer Family Foundation.

%
%

% End of text imported from the .//acknowledgements/Acknowledgements.tex input file

%
%
%
%
%
%
%

%
\clearpage
\printbibliography
\clearpage
\input{atlas_authlist}
%

%
%

%

%
%
%
%

\end{document}

%% file: atlas_authlist.tex
% ATLAS Collaboration author list
% Reference date of FTAG-2023-05 is 2025-03-26
% Author list last updated on date 17-JUL-25
% Data extracted on 17-Jul-2025 for paper reference FTAG-2023-05
% at 2:47pm
 
\begin{flushleft}
\hypersetup{urlcolor=black}
{\Large The ATLAS Collaboration}

\bigskip

\AtlasOrcid[0000-0002-6665-4934]{G.~Aad}$^\textrm{\scriptsize 104}$,
\AtlasOrcid[0000-0001-7616-1554]{E.~Aakvaag}$^\textrm{\scriptsize 17}$,
\AtlasOrcid[0000-0002-5888-2734]{B.~Abbott}$^\textrm{\scriptsize 123}$,
\AtlasOrcid[0000-0002-0287-5869]{S.~Abdelhameed}$^\textrm{\scriptsize 119a}$,
\AtlasOrcid[0000-0002-1002-1652]{K.~Abeling}$^\textrm{\scriptsize 55}$,
\AtlasOrcid[0000-0001-5763-2760]{N.J.~Abicht}$^\textrm{\scriptsize 49}$,
\AtlasOrcid[0000-0002-8496-9294]{S.H.~Abidi}$^\textrm{\scriptsize 30}$,
\AtlasOrcid[0009-0003-6578-220X]{M.~Aboelela}$^\textrm{\scriptsize 45}$,
\AtlasOrcid[0000-0002-9987-2292]{A.~Aboulhorma}$^\textrm{\scriptsize 36e}$,
\AtlasOrcid[0000-0001-5329-6640]{H.~Abramowicz}$^\textrm{\scriptsize 157}$
% ...
\bigskip
\\

% ...
$^{17}$Department for Physics and Technology, University of Bergen, Bergen; Norway.\\
% ...
$^{30}$Physics Department, Brookhaven National Laboratory, Upton NY; United States of America.\\
% ...
$^{36}$$^{(a)}$Facult\'e des Sciences Ain Chock, Universit\'e Hassan II de Casablanca;$^{(b)}$Facult\'{e} des Sciences, Universit\'{e} Ibn-Tofail, K\'{e}nitra;$^{(c)}$Facult\'e des Sciences Semlalia, Universit\'e Cadi Ayyad, LPHEA-Marrakech;$^{(d)}$LPMR, Facult\'e des Sciences, Universit\'e Mohamed Premier, Oujda;$^{(e)}$Facult\'e des sciences, Universit\'e Mohammed V, Rabat;$^{(f)}$Institute of Applied Physics, Mohammed VI Polytechnic University, Ben Guerir; Morocco.\\
% ...
$^{45}$Physics Department, Southern Methodist University, Dallas TX; United States of America.\\
% ...
$^{49}$Fakult\"{a}t Physik , Technische Universit{\"a}t Dortmund, Dortmund; Germany.\\
% ...
$^{55}$II. Physikalisches Institut, Georg-August-Universit\"{a}t G\"ottingen, G\"ottingen; Germany.\\
% ...
$^{104}$CPPM, Aix-Marseille Universit\'e, CNRS/IN2P3, Marseille; France.\\
% ...
$^{119}$$^{(a)}$New York University Abu Dhabi, Abu Dhabi;$^{(b)}$United Arab Emirates University, Al Ain; United Arab Emirates.\\
% ...
$^{123}$Homer L. Dodge Department of Physics and Astronomy, University of Oklahoma, Norman OK; United States of America.\\
% ...
$^{157}$Raymond and Beverly Sackler School of Physics and Astronomy, Tel Aviv University, Tel Aviv; Israel.\\
% ...
$^{a}$ Also at Affiliated with an institute formerly covered by a cooperation agreement with CERN.\\
$^{b}$ Also at An-Najah National University, Nablus; Palestine.\\
$^{c}$ Also at Borough of Manhattan Community College, City University of New York, New York NY; United States of America.\\
$^{d}$ Also at Center for Interdisciplinary Research and Innovation (CIRI-AUTH), Thessaloniki; Greece.\\
$^{e}$ Also at Centre of Physics of the Universities of Minho and Porto (CF-UM-UP); Portugal.\\
$^{f}$ Also at CERN, Geneva; Switzerland.\\
% ...

\end{flushleft}

% Created with Glance <Atlas.Glance@cern.ch>